\documentstyle[12pt,epsf]{article}
\begin{document}
\renewcommand{\theequation}{\arabic{equation}}
\textheight=19.3cm
\textwidth=12.3cm
\newcommand{\be}{\begin{equation}}
\newcommand{\ee}{\end{equation}}
\newcommand{\bdm}{\begin{displaymath}}
\newcommand{\edm}{\end{displaymath}}
\newcommand{\bea}{\begin{eqnarray}}
\newcommand{\eea}{\end{eqnarray}}
\newcommand{\co}{\; \; ,}
\newcommand{\nn}{\nonumber \\}
\newcommand{\hlogm}{\ln \frac{M^2}{\mu^2}}
\newcommand{\apbc}{(\alpha + \beta)^C}
\newcommand{\ambc}{(\alpha - \beta)^C}
\newcommand{\apbn}{(\alpha + \beta)^N}
\newcommand{\apbnc}{(\alpha + \beta)^{N,C}}
\newcommand{\apmbn}{(\alpha \pm \beta)^N}
\newcommand{\apmbcn}{(\alpha \pm \beta)^{C,N}}
\newcommand{\apmbnc}{(\alpha \pm \beta)^{N,C}}
\newcommand{\ambn}{(\alpha - \beta)^N}
\newcommand{\ambcn}{(\alpha - \beta)^{C,N}}
\newcommand{\apbcn}{(\alpha + \beta)^{C,N}}
\newcommand{\hmpp}{M_{\pi}}
\newcommand{\mppq}{M_{\pi}^2}
\newcommand{\p}[1]{(\ref{#1})}
\newcommand{\D}[1]{{\cal D}^{#1}}
\newcommand{\gag}{$\gamma \gamma \rightarrow \pi^0 \pi^0$}
\newcommand{\gpgpz}{$\gamma\pi^0 \rightarrow \gamma \pi^0$}
\newcommand{\gpgp}{$\gamma\pi \rightarrow \gamma \pi$}
\newcommand{\ggppz}{$\gamma\gamma\rightarrow\pi^0\pi^0\;$}
\newcommand{\ggppc}{$\gamma\gamma\rightarrow\pi^+\pi^-$}
\newcommand{\ggpp}{$\gamma\gamma\rightarrow\pi\pi\;$}
\newcommand{\mscript}[1]{{\mbox{\scriptsize #1}}}
\newcommand{\mtiny}[1]{{\mbox{\tiny #1}}}
\newcommand{\MS}{\mtiny{MS}}
\newcommand{\GeV}{\mbox{GeV}}
\newcommand{\MeV}{\mbox{MeV}}
\newcommand{\keV}{\mbox{keV}}
\newcommand{\ren}{\mtiny{ren}}
\newcommand{\kin}{\mtiny{kin}}
\newcommand{\hint}{\mtiny{int}}
\newcommand{\tot}{\mtiny{tot}}
\newcommand{\CHPT}{\mtiny{CHPT}}
\newcommand{\DISP}{\mtiny{DISP}}
\newcommand{\CA}{\mtiny{CA}}
\newcommand{\scs}{\co \;}
\newcommand{\sem}{ \; \; ; \;}
\newcommand{\per}{ \; .}
\newcommand{\la}{\langle}
\newcommand{\ra}{\rangle}
\newcommand{\bla}{\left\langle}
\newcommand{\bra}{\right\rangle}
\newcommand{\unith}{{\bf{\mbox{1}}}}
\renewcommand{\Sigma}{H}

\begin{titlepage}

\hspace{5cm} {}

\vskip-2cm

\begin{flushright}

BUTP--98/07\\

HIP--1998--10 /TH

\end{flushright}

\vskip2cm

\begin{center}

{{\Large{\bf{{Two--loop integrals in chiral perturbation

theory}}}}}

\vspace{1.5cm}

J. Gasser$^1$ and M.E. Sainio$^2$

\vspace{1.5cm}

{\small

$^{1}$ Institute of Theoretical Physics, University of Berne,
Sidlerstrasse 5, \\CH--3012 Berne, Switzerland

\vskip.5cm

$^{2}$
Dept. of Physics, University of Helsinki,
P.O. Box 9, FIN--00014 Helsinki,
Finland
 }

\vskip1cm

March 1998

\end{center}

\vskip1cm

\begin{abstract}
We consider chiral perturbation theory in the meson sector
at order $E^6$. In the terminology of the
external field technique,
the  two--loop graphs  so generated are  of the
sunset type.  We discuss the evaluation of several of these
in the case where the masses of the particles running in the loops are equal.
In particular, we present integral representations that are suitable for the
evaluation  of diagrams in kinematical regions where branch points and cuts
are present.
\end{abstract}

\end{titlepage}

\newpage

\setcounter{section}{0}
\setcounter{subsection}{0}

\section{Introduction\label{in}}
In the framework of chiral
perturbation theory (CHPT)
 \cite{chpt}, Green functions are expanded in
powers of the external momenta and of the light quark masses.
The generating functional is constructed by
use of an effective lagrangian and requires the
evaluation of tree graphs at leading order, one--loop graphs
at next--to--leading order, and two--loop graphs at
next--to--next--to--leading order. In this article, we
describe the evaluation of several  two--loop graphs in the
equal mass case.

The first  complete two--loop calculation in CHPT was per\-formed
in Ref. \cite{bgs},
in order to investigate the apparent discrepancy of the one--loop
prediction \cite{bico} of the cross section
$\gamma\gamma\rightarrow\pi^0\pi^0$
with the data \cite{crystal}. The topologies of
the two--loop graphs considered in \cite{bgs}
contain the ones in
\bea
\begin{array}{rl}
-&{\mbox{vector and axialvector two--point
functions}}\\
-&\pi\rightarrow e\nu\gamma\\
-&{\mbox{scalar  and vector form factors of the
pion}}\\
-& \pi\pi\rightarrow\pi\pi\\
-& \gamma\gamma\rightarrow\pi^+\pi^-\\
\end{array}
\label{eintrod1}\eea
In other words, knowing how to evaluate the two--loop graphs
in $\gamma\gamma\rightarrow\pi^0\pi^0$ allows one to calculate
those that occur in (\ref{eintrod1}) in the equal mass case. There are
other Green functions where the two--loop graphs have
the same topology as (1), e.g. $\pi\pi\rightarrow 4\pi$ or
$\gamma\rightarrow 4\pi$. The external momenta in those processes are,
however, in a different kinematical region than in (1), and the integral
representations worked out below do not apply.

The evaluation of the
two--loop integrals  in (1) is not
straightforward for several reasons:
 i) CHPT being a low--energy
expansion, one has to keep all masses at their physical
values -- the zero mass limit would result in a poor approximation
of the matrix element.
 ii) The interaction is of the
derivative type, which generates polynomials of high degree in
the numerator of the loop--integrals.
 iii)
In general, one needs the loop--functions in a region where branch
points and cuts are present.

Since the work of \cite{bgs}, additional two--loop calculations
have been performed. In the two flavour sector, these are the
amplitudes for $\gamma\gamma\rightarrow \pi^+\pi^-$ \cite{burgi},
$\pi\rightarrow e\nu\gamma$ \cite{bijnenstalavera}
and $\pi\pi\rightarrow\pi\pi$ \cite{bcges}. In the three flavour sector,
there exist calculations of the vector \cite{kambor1} and
axialvector \cite{kambor2} two--point
functions,  and of a combination of vector form
factors \cite{postschilcher}. For a review of these calculations, we refer the
reader
to \cite{bijnensmainz}.

The calculational methods developed  in \cite{bgs}
were  applied in \cite{burgi,bijnenstalavera,bcges,kambor2}.
 Further use of them is underway
 \cite{bijnensprivate}.
As these techniques  were never made public in a coherent manner,
we wish to do so here.
At the same time, we use the opportunity to simplify   the
originally used calculational tools.

We are, of course, aware that this is not the first publication on two--loop
integrals. Nevertheless, we feel that it would be inappropriate to give an
overview of what has been previously done in this field, because those
calculations are, as far as we can judge,
 mostly unrelated to what we aim at here. Indeed, in contrast to e.g. the
evaluation of two--loop integrals in the framework of the Standard Model,
where very different mass scales occur, the present article deals with
applications in $SU(2)\times SU(2)$ CHPT, where the masses are equal.
Furthermore, there is only a  limited number of two--loop graphs that
will ever need to be calculated. We expect that the techniques presented
below will be useful in this restricted framework, because they represent a
coherent method to deal with quite different topologies.
In addition,
the  same methods  can also be applied in chiral
$SU(3)\times SU(3)$ -- where the masses are different --
 see Ref.~\cite{kambor2} for the self--energy graph.
To give another illustration,
we expect that the two--loop graphs in $K_{l4}$
decays at order $E^6$ can be worked out with these methods in a
straightforward (yet admittedly tedious) manner.

The article is organized as follows. In section 2, we elucidate the structure
of the terms at order $E^6$ in the chiral expansion, in particular
the role of  the two--loop diagrams.  The following sections
are devoted to the
evaluation of the self--energy (section 3), the vertex (section 4), the
box (section 5) and the acnode diagram (section 6). Section 7 contains the
summary and concluding remarks. The notation is given in
appendix A, whereas appendix B contains
one--loop integrals. The divergences are evaluated and tabulated in appendix C.

\section{The diagrams at order $E^6$}
\begin{figure}[t]
\begin{center}
\mbox{\epsfysize=8cm \epsfbox{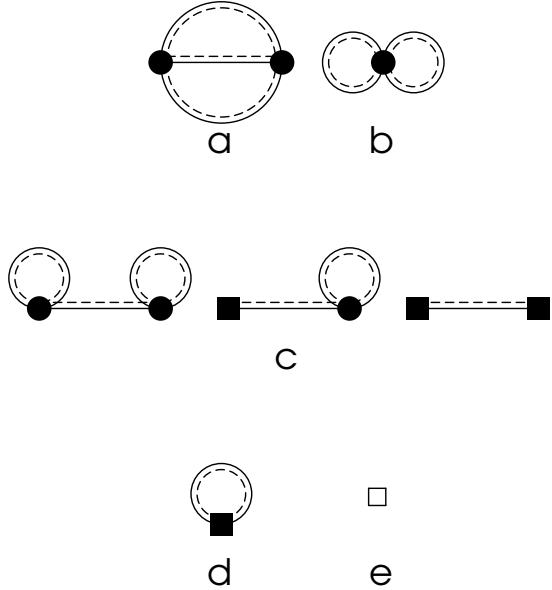}
     }
\caption{Contributions to the generating functional at order $E^6$. The
solid--dash lines denote the propagator in the presence of the external
fields. Filled circles (filled squares) denote vertices from the effective
lagrangians ${\cal L}_2$
(${\cal L}_4$) in Eq.~\protect{(\ref{esec21})}. The open square
in Fig.~e stands for  vertices
from ${\cal L}_6$. Only the sunset diagram Fig.~a generates
genuine two--loop integrals.}
\label{fz6} \end{center} \end{figure}

The effective lagrangian of QCD in the meson sector consists of
a string of terms,
         \bea
         {\cal L}_{\it eff} = {\cal L}_2 + \hbar {\cal
         L}_4 + \hbar^2 {\cal L}_6 + \cdots \scs\label{esec21}
         \eea
         where tree graphs with ${\cal L}_N$ generate
 polynomial contributions
of order $E^N$ in the energy  expansion.
These lagrangians contain
external sources  which allow one to evaluate  the
transition amplitudes with the background field method.
The path integral representation of the   generating
functional is
        \bea
         e^{i Z/\hbar} = \int [dU]
         e^{i/\hbar \int dx \, {\cal L}_{\it eff}}\co\nonumber
          \eea
where $[dU]$ denotes the chiral invariant measure.
 The
low--energy representation
\bdm
Z = Z_2 +  \hbar Z_4 + \hbar^2 Z_6 + \cdots
\edm
 is obtained by
expanding the lagrangians ${\cal L}_I$ around the
solution of the classical equation of motion
$\delta\! \int\! dx \, {\cal{L}}_2=0$ and carrying out the path integral to
the required order in $\hbar$. The diagrams which generate the terms of
order $E^6$ are collected   in $Z_6$ and displayed in
 figure~\ref{fz6}.
 The solid--dash lines
 stand for the propagator
 in the presence of the external fields. Full circles (full squares) denote
vertices from ${\cal L}_2$ (${\cal L}_4$), whereas the open square stands for
 a vertex from ${\cal L}_6$.
 The diagrams at order $E^6$
for a specific process
are obtained by attaching the external lines in all possible ways
to these graphs. Examples are the self--energy, the vertex, the box and
the acnode diagram considered below (figures
\ref{fself},\ref{fvertex},\ref{fbox} and \ref{facn},
respectively).

Figure~\ref{fz6}a
 collects all genuine two--loop diagrams. It is seen that, in the
language of the external field technique, all two--loop graphs
are of the sunset type. The figures~\ref{fz6}b-e display
diagrams that amount to products
of two one--loop integrals, to products of a one--loop integral
with a tree
graph contribution from ${\cal{L}}_4$, to one--loop graphs with ${\cal{L}}_4$,
or to tree graphs alone.
In the following, we reserve the term
"two--loop integral" to  contributions
from the sunset graph  Fig.~\ref{fz6}a.
[There
are Green functions where two--loop diagrams are
completely absent at order $E^6$ -- e.g., the vector two--point
functions
\cite{kambor1}. The evaluation of these matrix elements then
simplifies accordingly.]

In the following, we outline the evaluation
of the two--loop diagrams that occur in the process $\gamma
\gamma \rightarrow \pi^0\pi^0$ in
the two flavour case, with equal mass for the particles running in the loops.

\section{The self--energy}
\begin{figure}[h]
\begin{center}
\mbox{\epsfysize=5cm \epsfbox{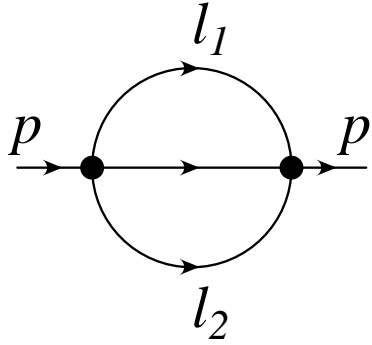}
     }
\caption{The self--energy diagram. The filled circles denote vertices
from the effective lagrangian ${\cal{L}}_2$ in Eq.~(2). The internal
lines stand for scalar propagators with mass $1$.}
\label{fself} \end{center} \end{figure}

We evaluate  contributions from the self--energy diagram
that is displayed in figure~\ref{fself}.
 The case where the masses of the particles running in the loops are not
identical is discussed e.g. in \cite{post} with a technique that is very
different from the one proposed in this work. References to earlier work on
the sunset graph may  be found in \cite{post}, see
also \cite{riesselmann}.

We consider the
integrals\footnote{The notation is given in appendix \ref{anotation}.}
\bea\label{seq20}
\left(\Sigma;
\Sigma^\mu;\Sigma^{\mu\nu}\right)=
\bla\bla\,(\,1;l_1^\mu;l_1^\mu
l_1^\nu\,)\,\prod_{i=1}^3\frac{1}{D_i} \bra\bra\co
\eea
with
\bea
D_1=1-l_1^2\scs D_2=1-l_2^2 \scs D_3=1-(p-l_1-l_2)^2\per
\eea
Integration over $l_2$ generates the loop--function
\bea
J(t)&=&C(w){\Gamma(-w)}\int_0^1dx\,[1-tx(1-x)]^w
\scs  t=(p-l_1)^2\per
  \eea
The function $J(t)$ is analytic in the complex $t-$plane, cut along the
positive real axis for $t\geq 4$. We insert the
Cauchy representation \cite{kacser}
\be\label{cauchy}
J(t)=\int_4^\infty\frac{[d\sigma]}{\sigma-t}\;\; ;\;\;
-1.5<w<0\co \ee
and integrate over
 $l_1$. In this manner, we obtain by use of the formulae in
appendix \ref{aoneloop}
  \bea\label{seq1}
\left(\Sigma;\Sigma^\mu\right)&=&
\int_4^\infty
[d\sigma]
\left\{F_2[z_2]\;;\;(1-x)F_2[z_2]\;p^\mu\right\}_1\co\nn
\Sigma^{\mu\nu}&=&
\frac{1}{2}\int_4^\infty
[d\sigma]
\left\{2(1-x)^2F_2[z_2]\;p^\mu p^\nu
-F_1[z_2]\;g^{\mu\nu}\right\}_1\co\nn
z_2&=&\left(1-s(1-x)\right)x +\sigma(1-x)\; ; \; s=p^2\per
\eea
The  integration over the variable $\sigma$
in Eq.~(\ref{seq1}) converges in the strip
\bdm
-1.5 < \mbox{Re} \; w < -1\per
\edm
Using partial integration in $x$ for the
last term
in $\Sigma^{\mu\nu}$, the nontrivial integrals in
Eq.~(\ref{seq1})  reduce to
\be
\label{seq30}
\int_4^\infty [d\sigma]\left\{ (1-x)^m F_2[z_2]\right\}_1
\per
 \ee

It remains to extract the finite and infinite parts in
Eq.~(\ref{seq30}) as $w\rightarrow 0$. We
subtract and add the
first two terms of the Taylor series of $F_2[z_2]$ around $s=1$.
The finite part   becomes
\bea
\int_4^\infty
d\sigma\beta\left\{(1-x)^m {\cal{K}}_2(x,\sigma;s)\right\}_1\co
\eea
where we have introduced the kernel
\bea
{\cal{K}}_2(x,\sigma;s)&=&
-\frac{1}{(16\pi^2)^2}\left\{\ln\frac{z_2}{z_2^{s=1}}
+(s-1)\frac{x(1-x)}{z_2^{s=1}}\right\}\per
\eea
It vanishes at $s=1$, together with its first derivative,
\bdm
{\cal{K}}_2(x,\sigma;1)={\cal{K}}'_2(x,\sigma;1)=0\co
\edm
and dies off rapidly at large values for $\sigma$,
\bdm
{\cal{K}}_2=O\left(\frac{1}{\sigma^2}\right)\co\;\sigma
\rightarrow \infty\per
\edm
 The  infinite part  of Eq.~(\ref{seq30}) may be
expressed in terms of the quantities
\bea
D(m,n)&=&\int_4^\infty
[d\sigma]\left\{(1-x)^mF_n[y]\right\}_1\co\nonumber\\
y&=&x^2+\sigma(1-x)\co
\eea
that are evaluated and tabulated in appendix \ref{adiv}.
As an illustration of the method, we consider  the scalar
integral $H(s)$. The subtracted function
\bdm
{\underline{\underline{H}}}(s)=H(s)-H(1)-(s-1)H'(1)
\edm
stays finite as $w\rightarrow 0$,
\be\label{seqm1}
{\underline{\underline{H}}}(s)=\int_4^\infty
d\sigma \beta\left\{{\cal{K}}_2(x,\sigma;s)\right\}_1\per
\ee
The poles at $w=0$ are contained in $H(1)$ and in $H'(1)$,
\bea
H(1)&=&D(0,2)\nn
&=&-C^2(w)\Gamma^2(-w)\left\{\frac{3}{2}
-\frac{17}{4}w+\frac{59}{8}w^2+O(w^3)\right\}\co\nn
 H'(1)&=&2\left\{D(1,3)-D(2,3)\right\}\nn
&=&-C^2(w)\Gamma^2(-w)\left\{\frac{1}{4}w+\frac{3}{8}w^2+O(w^3)
\right\}\per
\eea
The  representation (\ref{seqm1}) is well suited
for numerical evaluation at $s<9$ only.
  In the region $s>9$,
 $H(s)$ develops an imaginary part. At $d=4$,
\bea
\mbox{Im}\Sigma(s)=\frac{\pi}{s(16\pi^2)^2}\int_4^{(\sqrt{s}-1)^2}
d\sigma\beta[s-(\sqrt{\sigma}+1)^2]^{1/2}[s-(\sqrt{\sigma
} -1)^2]^{1/2}
\per
\eea
As a result, one has  the dispersion relation
\bea
{\underline{\underline{H}}}(s)=\frac{(s-1)^2}{\pi}\int_9^\infty
\frac{dz \, \mbox{Im} \Sigma(z)}{(z-1)^2(z-s)}\co
\eea
that allows one to evaluate ${\underline{\underline{H}}}$ also at
$s > 9$.  A similar remark applies to the  Lorentz
invariant components of the tensorial integrals $H^\mu$ and
$H^{\mu\nu}$.

\section{The vertex}
\subsection{Tensorial integrals}
\begin{figure}[t]
\begin{center}
\mbox{\epsfysize=5cm \epsfbox{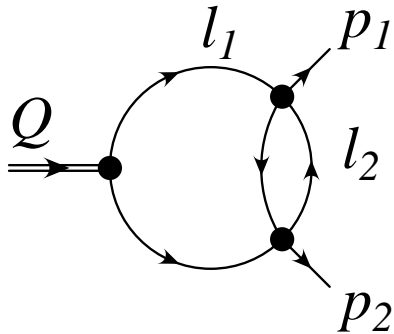}
     }
\caption{The vertex diagram. The filled circles denote
 vertices from the effective lagrangian ${\cal L}_2$ in Eq.~(2).
 The double external line
 denotes a current of momentum $Q$, e.g. one scalar or two electromagnetic
currents
 (contact term). The internal lines stand for scalar propagators with mass
$1$.}
\label{fvertex} \end{center} \end{figure}

Here we consider the vertex diagram Fig.~\ref{fvertex} that leads to the
integrals
\bea
\bla\bla\,(\,1\,;\,l_1^\mu\,;\,l_1^\mu
 l_1^\nu\,)\prod_{i=1}^4\frac{1}{D_i}\bra\bra \co\eea
with
\bea
D_1&=&1-l_1^2\scs D_2=1-(Q-l_1)^2\scs\nn
D_3&=&1-l_2^2\scs D_4=1-(l_2+l_1-p_1)^2\scs\nn
Q&=&p_1+p_2\scs p_1^2=p_2^2=1 \per
\eea
Integration over $l_2$ gives the loop--function $J(\bar{t})$
with $\bar{t}=(p_1-l_1)^2$, that we represent in the dispersive
manner  (\ref{cauchy}). We subtract the emerging subdivergence
 by writing $J(\bar{t})=J(0)+\bar{J}(\bar{t})$. The
contribution from $J(0)$
 generates a nonlocal divergence that is
removed by the usual renormalization procedure.
 We do
not consider this piece any further and
concentrate on the remainder,
  \bea
(V\,;\,V^\mu\,;\,V^{\mu\nu})&=&
\int_4^\infty
\frac{[d\sigma]}{\sigma}\bla\frac{\,(1\,;\,l_1^\mu\,;\,l_1^\mu
l_1^\nu)\,\bar{t}}{D_1D_2(\sigma-\bar{t})}\,
\bra\per
  \eea
We collect the denominators by
${\cal{F}}[D_1D_2(\sigma-\bar{t})]$ and rename the Feynman
parameters for later convenience,
\bdm
x_1,x_2\rightarrow x_2,x_3\co
\edm
such that
\bea\label{veq50}
(V\,;\,V^\mu\,;\,V^{\mu\nu})
&=&\int_4^\infty\frac{[d\sigma]}{\sigma}
\left\{\left\langle
\frac{(1; l_1^\mu;l_1^\mu l_1^\nu )\bar{t}}
{[z_3-(l_1-R)^2]^3}\right\rangle\right\}_{23}\co\nn
z_3&=&\sigma(1-x_3)+x_3^2y_2\scs
\nn y_2&=&1-sx_2(1-x_2)\scs\nn
 R&=&(1-x_2)x_3Q +(1-x_3)p_1\co\nn
s&=&Q^2=2p_1Q=2p_2Q\per
\eea
After the shift $l_1\rightarrow l_1+R$ the momentum integration
may be performed with (\ref{aonel0}), and the tensors
$V^\mu,V^{\mu\nu}$ can be expressed in terms of the scalar
integrals
\be\label{veq10}
V_m[P;s]=\int_4^\infty
\frac{[d\sigma]}{\sigma}\left\{P(x_2,x_3)
F_m[z_3]\right\}_{23}\co
 \ee
where $P(x_2,x_3)$ is a polynomial in $x_2,x_3$, and where the argument $s$ in
$V_m[P;s]$ denotes the $s$--dependence of $z_3$.
This
procedure automatically generates the tensorial structure in the
external momenta.
The integrals
(\ref{veq10}) can be decomposed into the convergent integral
$V_3[P;s]$ and the divergent polynomials $V_1[P;0]$ , $V_2[P;0]$
 by use of the recursion relation
\bea
V_m[P;s]&=&V_m[P_1(1,x_3);0]+mV_{m+1}[s[x_3^2(2x_2-1)]P_1;s]
 \co \nn
 P_1(x_2,x_3)&=&\int_0^{x_2}P(y,x_3)dy\co
\eea
 obtained from (\ref{veq10}) by partial
integration in $x_2$.
Finally,  $V_1[P;0]$ and $V_2[P;0]$ may be expressed
in terms of the  integrals
 \bea E(m,n)&=&\int_4^\infty
\frac{[d\sigma]}{\sigma}\left\{(1-x)^mF_n[y]\right\}_1
\co \nonumber\\
y&=&x^2+\sigma (1-x)\co
 \eea
that are evaluated and tabulated in appendix \ref{adiv}. In the following
subsection we illustrate the procedure in case of the scalar
integral $V(s)$.

\subsection{The scalar integral}

Performing the above described procedure, the scalar integral
$V(s)$ becomes
 \bea
V(s)  &=& V_3[P_s;s] -(w+2)\left\{E(0,2)-E(1,2)\right\}\co\nn
P_s&=&x_3^2\{y_2+ s(w+2)(1-2x_2)x_2\}\per
\eea
As $w\rightarrow 0$, the finite
part is
 \bea\label{veq12}
V_f(s)&=&\lim_{w\rightarrow 0}\, V_3[P_s;s]=\int_4^\infty
\frac{\beta d\sigma}{\sigma}v(s,\sigma)\co\nn
v(s,\sigma)&=&\frac{1}{(16\pi^2)^2}\int_0^1(1+sx_2(1-3x_2))dx_2
\int_0^1\frac{x_3^3
dx_3} {z_3}\per
\eea
 For \ggpp,
 this representation
 is not well suited, because
$V_f$ contains a branch point at $s=4$, and the physical
region for \ggpp is $s\geq 4$. This branch point manifests itself
in a zero in the denominator of the integrand in
Eq.~(\ref{veq12}) along
the curve $z_3=0$ in the square $0\leq x_2,x_3 \leq 1$. One may solve the
problem by writing a dispersion relation for $V_f$. Using
\bea
\frac{1}{z_3}=P(\frac{1}{z_3})+i\pi\delta(z_3)\label{eqdelta}
\eea
for $s \rightarrow s+i0^+$, we obtain
\bea
\mbox{Im} v(s,\sigma)&=&\frac{\pi}{(16\pi^2)^2}
\int_{x_{2_{-}}}^{x_{2_{+}}}
d x _ 2 x^3_{3_{+}}\frac{1+sx_2(1-3x_2)}{W_\sigma}\scs s>4\co\nn
x_{2_{\pm}}&=&\frac{1}{2}(1\pm(1-4/s)^{1/2})\co\nn
x_{3_{+}}&=&\frac{1}{2y_2}(\sigma-W_\sigma)\scs
W_\sigma=(\sigma^2-4\sigma y_2)^{1/2}\co\label{eqx2x3}
\eea
from where
\bdm
v(s,\sigma)=\frac{1}{\pi}\int_4^\infty \frac{dz}{z-s} \mbox{Im}
v(z,\sigma)\per
\edm
Integration over $\sigma$ gives
\bea\label{veq11}
V_f(s)=\frac{1}{\pi}\int_4^\infty\frac{dz}{z-s}
\int_4^\infty \frac{d\sigma} {\sigma}\,\beta\,
\mbox{Im} v(z,\sigma)\per
\eea
The function $V_f$ may be expressed in terms of
elementary functions \cite{bessis,bcges},
\bea
  {V}_f(s)=
 \frac{1}{(16 \pi^2)^2} \left[
\left( 3-\frac{\pi^2}{3 s \rho^2}\right) f +\frac{1}{2
\rho^2} f^2-
\frac{1}{3 s \rho^4} f^3  +\frac{25}{4} + \frac{\pi^2}{6} \right]
\;\;,
\eea
 with
\bea
f= \rho \left\{\ln\frac{1-\rho}{1+\rho} + i \pi
\right\}
\;\;; \; \; \rho=\sqrt{1-4/s}\;\;,\;\; s > 4\;\;.
\eea
Corresponding expressions hold for any $V_3[P;s]$.
In the evaluation of the matrix element for e.g. the process \ggpp,
 $V_3$ also occurs in the
 box diagrams considered below.
  Due to  the algebraic
complexity
of the expressions encountered,
 it may be more useful use
 representations analogous to Eq.~(\ref{veq11}).
 They allow for an efficient algebraic treatment.
The triple
integrals required are in any case considerably
easier to evaluate
than  the four--dimensional ones used in the
box  diagrams discussed below.

\section{The box}
\begin{figure}[t]
\begin{center}
\mbox{\epsfysize=5cm \epsfbox{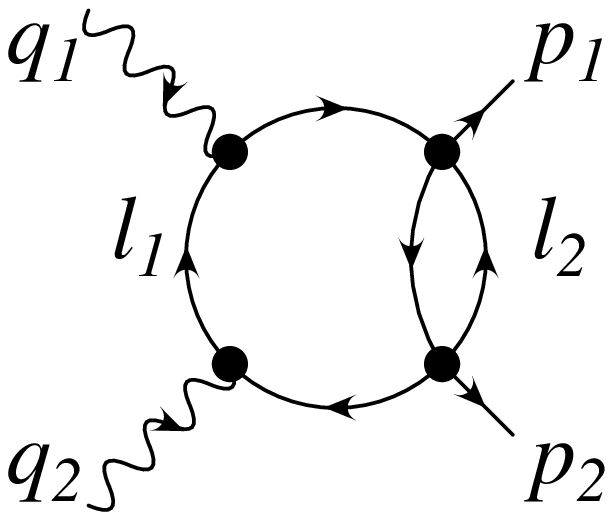}
     }
\caption{The box diagram. The filled circles
 denote vertices from the effective lagrangian ${\cal L}_2$ in Eq.~(2).
The wavy
lines stand for the electromagnetic current. The internal lines stand
for scalar propagators with mass $1$.}
\label{fbox} \end{center} \end{figure}

We consider integrals of the type
\bea
\bla\bla l_1^{\mu_1}\ldots
l_1^{\mu_N}\prod_{i=1}^{5}\frac{1}{D_i}\bra\bra\co
\eea
with
\bea
D_1&=&1-l_1^2\scs D_2=1-(l_1+q_1)^2 \scs\nn
D_3&=&1-(l_1-q_2)^2\scs\nn
D_4&=&1-l_2^2\scs D_5=1-(l_2+l_1+q_1-p_1)^2\per
\eea
These are generated by the diagram Fig.~\ref{fbox}.
We consider the case
\bdm
q_1+q_2=p_1+p_2\scs p_1^2=p_2^2=1\scs q_1^2=q_2^2=0\scs
\edm
that is relevant for the process \ggpp. Similarly to
the vertex diagram considered in the previous section,
integration over $l_2$ leads us to consider the tensors
\bea\label{beq1}
B^{\mu_1 \ldots \mu_N}&=&\int_4^\infty
\frac{[d\sigma]}{\sigma}\la l_1^{\mu_1} \ldots l_1^{\mu_N}
\frac{\bar{t}}{\sigma-\bar{t}}\prod_{i=1}^3\frac{1}{D_i}\ra\co\nn
\bar{t}&=&(p_1-q_1-l_1)^2\co
\eea
where we again have dropped the nonlocal singularity generated by
$J(0)$. The parametrization
${\cal{F}}[D_2D_1D_3(\sigma-\bar{t})]$ gives
\bea
B^{\mu_1 \ldots
\mu_N}=\int_4^\infty\frac{[d\sigma]}{\sigma}\left\{\bla
l_1^{\mu_1}\ldots
l_1^{\mu_N}\frac{\bar{t}}{[z_4-(l_1+\delta)^2]^4}\bra\right\}_{123}\co
\eea
with
\bea\label{beq7}
z_4&=&B-Ax_1\co\nn
A&=&x_2x_3\left\{s(1-x_2)x_3+(1-t)(1-x_3)\right\}
\equiv x_2x_3\bar{A}\co\nn
B&=&A+z_3=x_3^2+x_2x_3(1-x_3)(1-t)+\sigma(1-x_3)\co\nn
\delta&=&q_1x_1x_2x_3 -q_2x_3(1-x_2) +(q_1-p_1)(1-x_3)\co\nn
s&=&(p_1+p_2)^2,t=(p_1-q_1)^2\per
\eea
The quantity $z_3$  has already occurred in the vertex diagram, see
(\ref{veq50}).
With the shift $l\rightarrow l-\delta$, the momentum integrations are
easily done by use of Eq.~(\ref{aonel0}), and the tensors
$B^{\mu_1\ldots\mu_N}$ may be expressed in terms of the
scalar integrals
\bea\label{beq2}
B_m[P;s,t]&=&\int_4^\infty\frac{[d\sigma]}{\sigma}\left\{P(x_1,x_
2,x_3)F_m[z_4]\right\}_{123}\co
\eea
where $P(x_1,x_2,x_3)$ is a polynomial in $x_1,x_2,x_3$, and where the
arguments $s,t$ in $B_m[P;s,t]$ denote the $s,t$ dependence of $z_4$.
 These
integrals are convergent  at $w=0$ for $m\geq 3$.
 We reduce the divergent integrals to the case $m=3$ by
use of the recursion relation
\bea
B_m[P;s,t]&=&V_m[3x_2x_3P_1(1,x_2,x_3);s]-mB_{m+1}[AP_1(x_1,x_2,x_3);s,t]\co\nn
P_1(x_1,x_2,x_3)&=&\int_0^{x_1}dy\,P(y,x_2,x_3)\per
\eea
This relation is obtained from (\ref{beq2}) by partial
 integration
in $x_1$. The vertex functions $V_m$ have been discussed above,
and it remains to determine $B_{3,4}$.
In the kinematical region where $z_4\neq 0$,
 these functions may be obtained from
(\ref{beq2}) via a four--dimensional integration.
In the physical region for the process \ggpp, however, $z_4$ vanishes on a
two--dimensional surface embedded in the hypercube $0\leq
x_1,x_2,x_3 \leq 1$. Analogous singularities occur in the
physical region for $\gamma \pi \rightarrow \gamma \pi$. These
zeros in $z_4$ generate branch points at $s=4$ and at $t=9$.  We
therefore use again a Cauchy representation, and
 consider the region $t< 9$, where it suffices to use a
fixed--$t$ representation -- the region $t>9$ might then e.g. be
reached by use of a Mandelstam representation. We illustrate the procedure for
\bea\label{beq5}
G(s,t)&=&\lim_{w\rightarrow 0}B_3[P;s,t]= \int_4^\infty\frac{d\sigma}
{\sigma}\,\beta g(\sigma;s,t)\co\nn
g&=&\frac{1}{2(16\pi^2)^2}\left\{\frac{P(x_1,x_2,x_3)}{z_4}\right\}_{123}\per
\eea
By use of (\ref{eqdelta}), with $z_3\rightarrow z_4$, we obtain for the
discontinuity of $g$
\bea
\mbox{disc}_s g(\sigma; s,t)&\doteq& g(\sigma;
s+i0^+,t)-g(\sigma;s+i0^-,t)\nn
&=&\frac{6\pi i}{(16\pi^2)^2} \int_{x_{2_-}}^{x_{2_+}}
dx_2\int_{x_{3_+}}^1\frac{x_3}{\bar{A}} P\left(\frac{B}{A},x_2,x_3\right)
dx_3\co
\eea
where $x_{2_\pm}, x_{3_+}$ are given in (\ref{eqx2x3}). Therefore,
\bea\label{beq6}
G(s,t)=
\frac{1}{2\pi i}\int_4^\infty\frac{dz}{z-s}\int_4^\infty
\frac{d\sigma } {\sigma}\,\beta \, \mbox{disc}_s
g(\sigma;z,t)\per
\eea
In case that the
polynomial $P$ contains the
variable $s$, one
has to make sure to generate the correct asymptotic behaviour through the
dispersive representation. It may be necessary to pull out factors of $s$
 in the numerator before doing the dispersive integral -- it is in any case
useful to check the
 dispersive representation in a region free of cuts by use of
Eq.~(\ref{beq2}).
Analogous expressions can be obtained for $B_4$, e.g. by first
 integrating over $x_1$ and then again using Eq.~(\ref{eqdelta}).

\section{The acnode}
We consider the tensorial integral
\begin{figure}[t]
\begin{center}
\mbox{\epsfysize=6cm \epsfbox{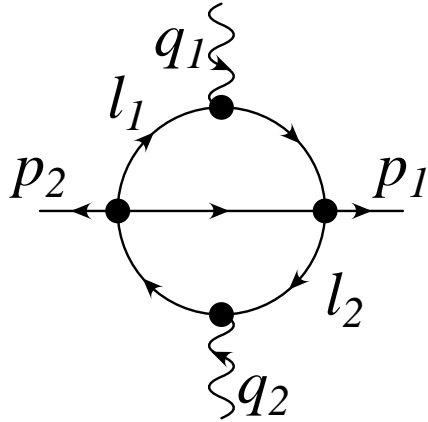}
     }
\caption{The acnode diagram. The filled circles
denote vertices from the effective lagrangian ${\cal L}_2$ in Eq.~(2).
The wavy
lines stand for the electromagnetic current. The internal lines stand
for scalar propagators with mass $1$.}
\label{facn} \end{center} \end{figure}
\bea\label{aeq0}
A^{\mu\nu}&=&\bla\bla\,{l_1^\mu}{
l_2^\nu}\,\prod_{i=1}^{5}\frac{1}{D_i}\bra\bra \co
\eea
that is generated by the acnode diagram Fig.~\ref{facn},
with
\bea
D_1&=&1-l_1^2\scs D_2=1-(l_1+q_1)^2 \scs D_3=1-l_2^2\co\nn
D_4&=&1-(l_2+q_2)^2\scs D_5=1-(l_2-l_1+p_1-q_1)^2\co\nn
q_1+q_2&=&p_1+p_2\co\nn
q_1^2&=&q_2^2=0,\; \; p_1^2=p_2^2=1\per
\eea
The first step in the calculation \cite{jetter} is to
employ the Feynman parametrization as
${\cal F}[D_1 D_2] \, {\cal F}[D_4 D_3]$ such that
\bea
A^{\mu\nu}&=&\int_0^1dx_1 \int_0^1dx_2\bla\bla
\frac{(l_1-(1-x_1)q_1)^\mu (l_2-x_2q_2)^\nu}
{(1-l_1^2)^2(1-l_2^2)^2(1-(l_1-l_2+k)^2)}\bra\bra\co\nonumber\\
k&=&x_1q_1+x_2q_2-p_1\per
\eea
The integrals $d^d l_1$ can now be performed once the denominators
 $(1-l_1^2)^2$ and $(1-(l_1-l_2+k)^2)$ are combined.
The result is proportional to
\bdm
\int^1_0 dx_3 \, x_3 \,
\frac{[(1-x_3)(l_2-k)^\mu-(1-x_1)\, q_1^\mu]}{[1-x_3(1-x_3)(l_2-k)^2]^
{1-w}}.
\edm
The  integrals  $d^d l_2$ can be done by combining
\bea
(1-l_2^2)^2([x_3(1-x_3)]^{-1}-(l_2-k)^2)^{1-w}\nonumber
\eea
in the standard manner.
The expression $A^{\mu \nu}$ then consists of a convergent part proportional
to tensors built from the external momenta, and a divergent piece proportional
 to $g^{\mu\nu}$. The convergent part is easy to evaluate numerically in the
physical region for
  $\gamma\gamma\rightarrow\pi\pi$, because it does not contain branch points
there. In case one wishes to evaluate these terms e.g. at
$(q_1-p_1)^2 \geq 9$,
one may again write a dispersion relation for the form factors in question.
Here, we concentrate on the term
 proportional to $g^{\mu\nu}$,
\bea
A^{\mu\nu}&=&A_0g^{\mu\nu}+\cdots\; ,\nonumber\\
A_0&=& -C^2(w)\Gamma(-2w)\frac{1}{2} \int_0^1 d^4x
 [x_3(1-x_3)]^{-w} x_4 (1-x_4)^{-w} \,{z_5}^
{2 w}\; ,\nn
\eea
where
\bea
{z_5} = 1-x_4 +x_3(1-x_3) x_4\left\{1-(1-x_4)k^2\right\}\per
\eea
Expanding the integrand in powers of $w$, we find
\bea
A_0&=&C(w)^2 \Gamma(-w)^2\frac{w}{16}(2+7w)
+\frac{1}{2(16\pi^2)^2}\int_0^1 d^4x \;x_4 \ln{z_5} + O(w)
\per\nn
\eea

\section{Summary and conclusions}
\begin{enumerate}
\item
We have discussed in this article the two--loop
diagrams that occur in the evaluation of the amplitudes
 $\gamma
\gamma\rightarrow \pi^0\pi^0$
in the equal mass case. These are the self--energy, the vertex,
 the box
and the
 acnode  graphs, displayed in Figs.~\ref{fself}--\ref{facn}.  These
two--loop graphs are also the
ones that occur in the
processes listed in Eq.~(\ref{eintrod1}), and no less.
 \item
We first discuss our results for the {\it{self--energy}}, the {\it{vertex}}
and the
{\it{box graphs}}.
 These diagrams contain, as a subgraph,
the one--loop
function\footnote{For the notation, see appendix \ref{anotation}.}
\bea
J(t)=\left\la\frac{1}{(1-l_1^2)}\frac{1}{(1-(l_1-p)^2}
\right\ra \;\; ; \; t=p^2\scs\nonumber
\eea
that we represent  in a dispersive manner,
\bea
J(t)=\int_4^\infty \frac{[d\sigma]}{\sigma-t}
\per\nonumber
\eea
As a result, they
 may be represented in  $d$ dimensions
 as linear combinations of the
following integrals:
\bea
\Gamma(-w-n)\int_4^\infty
[d\sigma]\{P_1z_2^{w+n}\}_1
&&\hspace{2cm}\mbox{Fig.}~\ref{fself}\;({\mbox{self--energy}})\nonumber\\
&&\nonumber\\
\Gamma(-w-n)\int_4^\infty
\frac{[d\sigma]}{\sigma}\left\{P_2z_3^{w+n}\right\}_{23}
&&\hspace{2cm}\mbox{Fig.}~\ref{fvertex}\;({\mbox{vertex}})\nonumber\\
&&\nonumber\\
\Gamma(-w-n)\int_4^\infty
\frac{[d\sigma]}{\sigma}\left\{P_3z_4^{w+n}
\right\}_{123}
&&\hspace{2cm}\mbox{Fig.}~\ref{fbox}\;({\mbox{box}})\nonumber\\
\nonumber
\eea
Here, $n$ denotes an integer, and $P_k$ are
polynomials in
 $k$ Feynman parameters. The  $z_i$ are polynomials
in the external
momenta, in the Feynman parameters  and in $\sigma$.
\item
The vertex and box integrals receive fur\-ther contri\-bu\-tions,
that  con\-tain a  nonlocal singularity which is generated by a
divergent
subdiagram. These contributions are cancelled by the standard
renormalization procedure, and we have not considered them
further here.
 \item
In order to recover the finite and infinite parts in the vertex and in
the box diagrams at $w\rightarrow 0$,
 we have performed partial integrations in the
Feynman parameters, reducing in this manner the exponent in
$z_i^{w+n}$.
The finite parts are obtained by
reducing the exponent to $n=-1$,
 while the surface terms generated by  partial integration
produce  the  divergences.
 In the case of the self--energy
diagram, one does not  obtain a finite
result in this manner. We have instead subtracted the first two
terms of the Taylor series expansion of $z_2^w$ around $s=1$.
\item
In this manner, we are able to express all divergences in terms of the
integrals
\bea
\left\{D(m,n)\; ; \; E(m,n)\right\}=\int_4^\infty[d\sigma]
\left\{ 1 \; ; \; \frac{1}{\sigma}\right\} \,
\left\{ (1-x)^mF_n[y]\right\}_1\co\nonumber
\eea
where
\bdm
 y=x^2 +\sigma (1-x)\; ; \; m=0,1,2\ldots; \; \; n=1,2,3\ldots
\edm
We evaluate and tabulate these quantities in appendix \ref{adiv}.
\item
In the physical region for
 $\gamma
\gamma\rightarrow \pi^0\pi^0$, the vertex and box
diagram develop branch points and cuts, as a result of which the above
representation for the finite part is not appropriate.
 We instead write
fixed--$t$ dispersion relations. We provide an integral representation for
the required absorptive part in each case.
\item
The vertex diagrams can be given in closed form \cite{bessis,bcges}.
The algebraic complexities in the case of $\gamma\gamma\rightarrow\pi\pi$
suggest, however,  that it may sometimes be simpler to keep
them in the form of the integral representations
provided here. Needless to say that this decision is a matter of taste.
\item
Finally, we come to the {\it{acnode diagram}}, shown in  Fig.~\ref{facn}.
Instead of presenting the originally used \cite{bgs,burgi} method,
we evaluate it here along lines that are similar to the ones suggested in
Ref.~\cite{jetter} for the decay $\eta\rightarrow \pi^0\gamma\gamma$.
The kinematics in \ggppz allows for a substantially simpler procedure
than the one needed in \cite{jetter}.
\item
We conclude that, with these methods at hand, one is able to
calculate
many processes at two--loop order in the framework of
chiral perturbation theory.
\end{enumerate}

\section*{Acknowledgements}
We thank
S. Bellucci, J. Bijnens, U. B\"urgi, G. Colangelo, G. Ecker and P. Talavera
for enjoyable discussions and collaboration at various stages
of this work.
In particular, we thank J. Bijnens, G. Colangelo and P. Talavera for checking
our result for
 the self--energy and for the vertex diagrams, and for useful
comments on the manuscript. One of us (JG) has been partially supported by
the Swiss National Science Foundation. He furthermore acknowledges
the warm
hospitality of the University of Granada, where this work was completed.
The other author (MES) thanks
Oskar \"Oflunds Stiftelse for a travel grant.
\newpage

\newcounter{zahler}
\renewcommand{\thesection}{\Alph{zahler}}
\renewcommand{\theequation}{\Alph{zahler}\arabic{equation}}

\setcounter{zahler}{0}
\appendix

\setcounter{equation}{0}
\addtocounter{zahler}{1}
\renewcommand{\thesection}{\Alph{zahler}}
\renewcommand{\theequation}{\Alph{zahler}\arabic{equation}}

\section{Notation }
\label{anotation}
 To simplify the notation, we set the
pion masses equal to one,
\bdm
M_{\pi^\pm}=M_{\pi^0}=1\per
\edm
As is customary, we use dimensional regularization and put
\bdm
w=\frac{d}{2}-2\co
\edm
where $d$ denotes the dimension of space-time.
Loop integrations are symbolized by a bracket,
\bea
\la\ldots\ra &=& \int \frac{d^dl_1}{i(2\pi)^d}(\ldots)\co\nn
\la\la\ldots\ra\ra&=&\int\frac{d^dl_1}{i(2\pi)^d}\int\frac{d^dl_2
} {i(2\pi)^d}(\ldots)\per
\eea
We combine denominators with
\bea
[a_1\ldots a_N]^{-1}&=&\int
[dx]_{N-1}[a_1x_1\ldots
 x_{N-1}+a_2x_2\ldots x_{N-1}(1-x_1)+\nn
&&a_3 x_3\ldots x_{N-1}(1-x_2)+\ldots +a_N(1-x_{N-1})]^{-N}\nn
&\equiv& {\cal{F}}[a_1\ldots a_N].
\eea
Here $[dx]_N$ stands for the normalized measure
\bea
[dx]_N&=&N!\prod_{\nu=1}^N\theta[x_\nu(1-x_\nu)]x_\nu^{\nu-1}
dx_\nu\co\nn
\int[dx]_N&=&1\co
\eea
and $\theta(x)$ denotes the step function.
 We abbreviate multiple
Feynman integrals by
\bea
\{\ldots\}_1&=&\int_0^1 dx\{\ldots\} \scs\nonumber\\
\{\ldots\}_{23}&=&2\int_0^1 dx_2\int_0^1 x_3 dx_3\{\ldots\} \scs\nonumber\\
\{\ldots\}_{123}&=&6\int_0^1 dx_1\int_0^1 x_2 dx_2
\int_0^1 x_3^2dx_3\{\ldots\} \per
 \eea
Furthermore,
 we use the measure
\bea
[d\sigma]&=&\frac{C(w)\Gamma(3/2)}
{\Gamma(3/2 + w)}
 (\frac{\sigma}{4}-1)^w\beta \;d\sigma\co
\eea
with
\bea
C(w)=\frac{1}{(4\pi)^{2+w}}\scs \; \; \;
\beta=(1-4/\sigma)^{1/2}
\co
\eea
and
\bea
\lim_{w\rightarrow
0}\;[d\sigma]=\frac{\beta}{16\pi^2}\;d\sigma\per \eea

\setcounter{equation}{0}
\addtocounter{zahler}{1}

\section{One--loop integrals }
\label{aoneloop}
In the text we use the loop--functions
\bea\label{aonel0}
\bla\frac{1}{[z-l_1^2]^m}\bra &=&F_m[z]\co\nn
\bla\frac{l_1^\mu l_1^\nu}{[z-l_1^2]^m}\bra
&=&-\frac{g^{\mu\nu}}{2(m-1)}\;F_{m-1}[z]\co\nn
\bla\frac{l_1^\mu l_1^\nu l_1^\rho\l_1^\sigma}{[z-l_1^2]^m}\bra
&=&\frac{g^{\mu\nu}g^{\rho\sigma}+\mbox{cycl.}}{4(m-1)(m-2)}
\;F_{m-2}[z] \per
\eea
They are  given by
\bea
F_m[z]=z^{w+2-m}C(w)\frac{\Gamma(m-2-w)}{\Gamma(m)}\co\; m\geq
1\per \eea
In particular,
\bea
\left(F_1\,;\,F_2\,;\,F_3\,;\,F_4\right)&=&z^w\,C(w)\times\nn
&&\times\left(
\Gamma(-1-w)z \,;\, \Gamma(-w)\,;\,
\frac{\Gamma(1-w)}{2z}\,;
\,\frac{\Gamma(2-w)}{6z^2}\right)\per\nn
\eea
We also use
\bea
J(t)=\bla\frac{1}{1-l_1^2}\frac{1}{1-(l_1-p)^2}\bra\sem t=p^2\co
\eea
with
\bea
J(0)=C(w)\Gamma(-w)\per
\eea

\setcounter{equation}{0}
\addtocounter{zahler}{1}

\section{The integrals $D(m,n)$ and $E(m,n)$}
\label{adiv}
Here we consider the  integrals
\bea
\left\{D(m,n)\; ; \; E(m,n)\right\}=\int_4^\infty[d\sigma]
\left\{ 1 \; ; \; \frac{1}{\sigma}\right\} \,
\int_0^1dx\,(1-x)^mF_n[y]
\eea
where
\bdm
 y=x^2 +\sigma (1-x)\; ; \; m=0,1,2\ldots\; ,\;n=1,2,3\ldots
\edm
In particular, we determine the divergent parts in
$D(m,n\leq 3)$ and in $E(m,n\leq 2)$.
By partial integration in $x$, we obtain
 the recursion relation
\bea
(3\!+\!w\!+\!m\!-\!n)D(m,n)&=&\frac{\Gamma(n\!-\!w\!-\!2)Q(w+2-n)}{\Gamma(n)
\Gamma(-w)} \nn
 &&-n\left\{D(m,n+1)\!-\!D(m+2,n+1)\right\}\co\label{recurs1}\nn
\eea
with
\bea
Q(\alpha)&=&C(w)\Gamma(-w)\int_4^\infty [d\sigma]
\; \sigma^\alpha\nn
&=&
C^2(w)\Gamma(-w)\Gamma(-1-w-\alpha)\frac{\Gamma(-\alpha)}{\Gamma(-2\alpha)}\per
\eea
An analogous relation holds for $E(m,n)$ ,
 with $Q(w+2-n)\rightarrow Q(w+1-n)$.
 One may
use these recursion relations to express $D(m\geq 1,n\leq 3)$ and
$E(m,n\leq 2)$ through the
divergent quantities $Q$ and the convergent integrals $D(m\geq 1,4)$
and $E(m,3)$. The case $D(0,n)$
must be treated separately, see below.

\subsection{Explicit expressions for $D(m,n)$}
Let
\bea
D(m,n)&\hspace{-3mm}=&\hspace{-3mm}
C^2(w)\Gamma^2(-w)\left\{p(m,n,0)\!+\!wp(m,n,1)
\!+\! w^2 p(m,n,2) \!+\! O(w^3)\right\}\per\nonumber\\
\eea
For $m\geq 1$, we proceed as described above and find
\newcommand{\cs}{, \;}
\begin{eqnarray}
p(m,1,0)&=&(m^2+4m+5)m(m+4)N_4\cs\nonumber\\
p(m,1,1)&=&-(2m^7+29m^6+172m^5+540m^4
+964m^3\nonumber\\
&&+951m^2+430m+36)m(m+4)N_4^2\cs\nonumber\\
p(m,1,2)&=&\left\{m(m+1)D(m+6)-3m(m+3)D(m+4)\right.\nonumber\\
 &&+\left.3(m+1)(m+4)D(m+2)-(m+3)(m+4)D(m)\right\}N_4\nonumber\\
&&+\left(3m^{12}+74m^{11}+812m^{10}+5230m^9+21938m^8
\right.\nonumber\\
&&+\left.62724m^7+
123986m^6+167682m^5+149409m^4+81146m^3
\right.\nonumber\\
&&+\left.23372m^2+3456m+864\right)m(m+4)N_4^3\cs\nonumber\\
p(m,2,0)&=&-m(m+2)N_2\cs\nonumber\\
p(m,2,1)&=&(2m^3+7m^2+7m+1)m(m+2)N_2^2
\cs\nonumber\\
p(m,2,2)&=&\left\{mD(m+4)-2(m+1)D(m+2)+(m+2)D(m)\right\}N_2\nonumber\\
&&\hspace{-.6cm}-\left(4m^6+26m^5+61m^4+63m^3+27m^2+4m+2\right)m(m+2)N_2^3
\nonumber\\
p(m,3,0)&=&0\cs\nonumber\\
p(m,3,1)&=&-1/(4m)\cs\nonumber\\
p(m,3,2)&=&\left\{D(m+2)-D(m)\right\}/(2m)-(2m-1)/(4m^2)\cs\nonumber\\
\end{eqnarray}
where
\begin{eqnarray}
N_2^{-1}&=&m(m+1)(m+2)\cs N_4^{-1}=N_2^{-1}(m+3)(m+4)\cs\nonumber\\
D(m)&=&\int_4^\infty d\sigma\beta\int_0^
1 \frac{dx(1-x)^m}{\left\{x^2+\sigma(1-x)\right\}^2}\nonumber\\
&=&\int_0^1 dx \frac{x^{m-1}}{(1-x)^2}
\left( 1+\frac{2x}{1-x^2}\ln{x}\right)\per
\end{eqnarray}
For example,
\bea
D(1)&=&(\pi^2-4)/16\cs D(2)=-(\pi^2-12)/16\cs D(3)
=-(13\pi^2-132)/48\per\nn
\eea
In table 1, we display some of the coefficients $p(m,n,k)$ for convenience.
We now turn to  $D(0,n)$ which is divergent for any $n$. The
recursion relation (\ref{recurs1}) allows one to evaluate
$D(0,1)$ and $D(0,3)$ from
\bea
D(0,2)=C(w)\Gamma(-w)\int_4^\infty[d\sigma]
\,
\int_0^1dx\,\left(x^2+\sigma (1-x)\right)^w\scs\label{ed02}
\eea
and from $D(m\geq 2,n)$. In order to evaluate $D(0,2)$,
we add and subtract from the integrand in (\ref{ed02}) the quantity
\bea
\triangle=\left( x+\sigma (1-x)\right)^w -\frac{wx}{\sigma}
\left(\sigma (1-x)\right)^w
\per\nonumber
\eea
The integral
\bea
C(w)\Gamma(-w)\int_4^\infty[d\sigma]
\,
\int_0^1dx\,\left((x^2+\sigma(1-x))^w - \triangle\right)
\eea
is finite at $w=0$, whereas the divergence is contained in
\bea
C(w)\Gamma(-w)\int_4^\infty[d\sigma]
\,
\int_0^1dx\,\triangle\per
\eea
In this manner, we obtain the values $p(0,n\leq 3,k)$ displayed in table 1.

\subsection{Explicit expressions for $E(m,n)$}
Let
\bea
E(m,n)&\hspace{-3mm}=&\hspace{-3mm}
C^2(w)\Gamma^2(-w)\left\{q(m,n,0)\!+\!wq(m,n,1)
\!+\! w^2 q(m,n,2) \!+\! O(w^3)\right\}\per\nonumber\\
\eea
Proceeding in the manner described above, we find
\begin{eqnarray}
q(m,1,0)&=&(m^2+4m+2)N_3\cs\nonumber\\
q(m,1,1)&=&-(3m^5+31m^4+124m^3+235m^2+205m+64)N_3^2
\cs\nonumber\\
q(m,1,2)&=&\left\{(m+1)E(m+4)\!-\!2(m+2)E(m+2)
\!+\!(m+3)E(m)\right\}N_3\nonumber\\
&&+\left(7m^8+115m^7+802m^6+3097m^5+7230m^4\nonumber\right.\\
&&+\left.10425m^3
+9041m^2+4295m+848\right)N_3^3\cs\nonumber\\
q(m,2,0)&=&N_1/2\cs\nonumber\\
q(m,2,1)&=&(2m+1)N_1^2/2\cs\nonumber\\
q(m,2,2)&=&\left\{E(m+2)-E(m)\right\}N_1
+(4m^2+6m+3)N_1^3/2\cs\nonumber\\
\end{eqnarray}
where
\begin{eqnarray}
N_1^{-1}&=&(m+1)\cs
N_3^{-1}=N_1^{-1}(m+2)(m+3)\cs\nonumber\\
E(m)&=&\int_4^\infty\frac{d\sigma}{\sigma}\beta\int_0^
1 \frac{dx(1-x)^m}{x^2+\sigma(1-x)}\nonumber\\
&=&-\int_0^1dx\frac{x^m}{(1-x)^2}
\left( 2+\frac{1+x}{1-x}\ln{x}\right)\per
\end{eqnarray}
For example
\begin{eqnarray}
E(0)&=&\frac{1}{2}\cs
E(1)=\frac{\pi^2-9}{6}\cs
E(2)=\frac{4\pi^2-39}{6}\per
\end{eqnarray}
For convenience, we display some of the coefficients $q(m,n,k)$ in table 2.
 \begin{table}[p]
\begin{center}
\caption{The coefficients $p(m,n,k)$.\label{tpikm}
         }
\vspace{1em}
\begin{tabular}{ccccc} \hline
&&&&\\
$m$ &$n$&$p(m,n,0)$&$p(m,n,1)$&$p(m,n,2)$\\
&&&&\\
$0$&$1$&$\frac{13}{12}$&$-\frac{469}{144}$&$
\frac{10445}{1728}$\\
&&&&\\
$1$&$1$&$\frac{5}{12}$&$-\frac{781}{720}$&$
\frac{78121}{43200}$\\
&&&&\\
$2$&$1$&$\frac{17}{60}$&$-\frac{2389}{3600}$&$
\frac{233857}{216000}$\\
&&&&\\
$0$&$2$&$-\frac{3}{2}      $&$\frac{17}{4}
$&$-\frac{59}{8} $\\
&&&&\\
$ 1$&$2$&$-\frac{1}{2}      $&$\frac{17}{12}$
&$-\frac{59}{24} $\\
&&&&\\
$ 2$&$2$&$-\frac{1}{3}$&$\frac{59}{72}$&$-\frac{1333}{864}$\\
&&&&\\
$ 0$&$3$&$\frac{1}{4}$&$-\frac{1}{2}$&$-\frac{\pi^2}{6}+1$\\
&&&&\\
1&3&0&$-\frac{1}{4}$&$-\frac{\pi^2}{6}+\frac{5}{4}$\\
&&&&\\
2&3&0&$-\frac{1}{8}$&$-\frac{\pi^2}{6}+\frac{23}{16}$\\
&&&&\\
3&3&0&$-\frac{1}{12}$&$-\frac{\pi^2}{6}+\frac{3}{2}$\\
&&&&\\
4&3&0&$-\frac{1}{16}$&$-\frac{\pi^2}{6}+\frac{883}{576}$\\
&&&&\\ \hline
\end{tabular}
\end{center}
\end{table}

 \begin{table}[p]
\begin{center}
\caption{The coefficients $q(m,n,k)$.\label{tqikm}
         }
\vspace{1em}
\begin{tabular}{ccccc} \hline
&&&&\\
$m$ &$n$&$q(m,n,0)$&$q(m,n,1)$&$q(m,n,2)$\\
&&&&\\
0&1&$\frac{1}{3}$&$-\frac{16}{9}$&$\frac{223}{54}$\\
&&&&\\
1&1&$\frac{7}{24}$&$-\frac{331}{288}$&$\frac{9011}{3456}$\\
&&&&\\
2&1&$\frac{7}{30}$&$-\frac{1499}{1800}$&$\frac{206087}{108000}$\\
&&&&\\
0&2&$\frac{1}{2}$&$ \frac{1}{2}$&$\frac{2\pi^2}{3}-\frac{11}{2}$\\
&&&&\\
1&2&$\frac{1}{4}$&$\frac{3}{8}$&$\frac{2\pi^2}{3}-\frac{93}{16}$\\
&&&&\\
2&2&$\frac{1}{6}$&$\frac{5}{18}$&$\frac{2\pi^2}{3}-\frac{325}{54}$\\
&&&&\\
3&2&$\frac{1}{8}$&$\frac{7}{32}$&$\frac{2\pi^2}{3}-\frac{7073}{1152}$\\
&&&&\\
4&2&$\frac{1}{10}$&$\frac{9}{50}$&$\frac{2\pi^2}{3}-\frac{27983}{4500}$\\
&&&&\\
\hline
\end{tabular}
\end{center}
\end{table}

\end{document}